\def\beg{\begin{equation}}
\def\eeq{\end{equation}}

\documentstyle[12pt]{article}
\begin{document}
\baselineskip18pt
\begin{center}
{\Large{\bf  Interaction and magnetic moment of the electron in quantum Hall effect}}
\vskip0.35cm
{\bf Keshav N. Shrivastava}\\
{\it School of Physics, University of Hyderabad, \\
Hyderabad  500046, India.}
\end{center}
\vskip0.5cm
We introduce $\hat{s}.\hat{n}$ where $\hat{n}=\vec{r}/|\vec{r}|$
which along with the velocity forms a spin-orbit interaction of the
order of $v/c$ whereas the usual spin-orbit interaction is of the 
order of $v^2/c^2$. The effective fractional charges obtained from
this interaction are in good agreement with the experimental data of
quantum Hall effect. The energy of the system diverges as $l\to\infty$
indicating that there is a phase transformation at the half filled Landau
level.
\newpage
\section{Introduction}

In 1876, Edwin Hall had found that the resistance along the $y$
direction varies linearly with the magnetic field in a conductor
when voltage is applied along the $x$-direction and the magnetic
field along the $z$ direction. When the applied magnetic field
is very large such as 5 Tesla, von Klitzing et al [1] found that
a plateau appears in the Hall resistivity at $h/\nu e^2$ where
$\nu$ is an integer. The value of $h/e^2$ can be determined
very accurately at $\nu=1$. For still larger magnetic fields,
fractional values of $\nu$ such as 1/3 were found by Tsui,
St\"ormer and Gossard [2]. By assuming that the charge of the
quasiparticles is (1/3)$e$, Laughlin [3] has written a wave
function for the fractionally charged quasiparticles. This wave
function is correct as far as the quantum field theory is
concerned except that it does not predict the charges. It
implies that the repulsive Coulomb interactions by themselves
will produce a quasiparticle of charge (1/3)$e$. However, our
theory suggests [4] that the charge of (1/3)$e$ cannot be
produced unless the spin is considered. The question is that
whether our theory is a small correction to Laughlin's? The
answer is that it is not a correction to Laughlin's theory. It
is all together different. We obtained [4] the series of
fractional charges, by using an effective gyromagnetic ratio,
which have Kramers particle-hole symmetry [5]. Since the Bohr
magneton is involved our effective gyromagnetic ratios can be
interpreted as fractional charges. Our theory is also described
in a book [6]. There is no doubt that our values of the
effective fractional charges are in full agreement with
experimental data [7-11]. Since the agreement is very good, it
is necessary to understand the underlying principles.

In this paper, we explain the interaction which gives the
correct series of fractional charges and show that there is a
new correction to the Bohr magneton at a very large, $\sim8 - 20$
Tesla, magnetic field. At $l\to\infty$, there is a divergence in
the energy showing that there is a phase transition associated
with symmetry breaking. 

\section{Pseudoscalar}

From the invariance of the hamiltonian under noncontinuous
orthogonal transformation of coordinates represented by the
reflection, $r\to-r$, it may be concluded that parity is
constant. If the hamiltonian is invariant for $r\to-r$, then
parity is conserved. Because this transformation can not be
generated by continuous rotation, the parity is independent of
angular momentum. If ${\cal H}$ is invariant under the substitution,
\begin{eqnarray}
\phi(r,t) &\to& \phi(-r,t)\nonumber\\
P_+ H P_+^{-1} &=& H\nonumber\\
{}[P_+,H]&=&0
\end{eqnarray}
or $P_+$ is a constant. If both $H$ and the commutation
relations are also invariant under $\phi(r,t)\to-\phi(-r,t)$,
one can define a reflection,
\begin{eqnarray}
P_-\phi(r,t)P_-^{-1}&=&-\phi(-r,t)\nonumber\\
\mbox{and}\qquad\qquad\qquad P_-&=&\mbox{constant.}
\end{eqnarray}
Which of the two operators $P_\pm$ is a constant can be
determined from the known interaction. If ${\cal H}$ includes a term
$\int\rho(r)\phi(r,t)d^3r$, where $\rho(r)$ is invariant under
reflection then only $P_+$ commutes with ${\cal H}$ and $\phi$ is then
a scalar. The term $\int d^3rs.\nabla\phi(r,t)$ with $P_\pm s
P_\pm^{-1}=s$ commutes only with $P_-$ and hence $\phi$ is
called a pseudoscalar,
\begin{eqnarray}
{}[P_-, \rho(r) s.\nabla \phi(r,t)] = 0\,\,.
\end{eqnarray}
Substituting $\vec{\nabla}=m\vec{v}/(-i\hbar)$, we can write
\begin{eqnarray}
{}[P_-, \rho(r) s.v \phi(r,t)] = 0\,\,,
\end{eqnarray}
for the pseudoscalar. We will now construct a {\it true scalar}
which gives the correct fractional charges, exactly as observed
in the experiments. In a non-relativistic theory, the
interaction of the electron with the self-consistent field is
independent of the spin. Such a dependance can be introduced by
a term proportional to $s.\hat{n}$ where $\hat{n}$ is a unit
vector in the direction of the radius vector $\vec{r}$ of the
particle, $\hat{n}=\vec{r}/|\vec{r}|$ and thus product is a
pseudoscalar. The dependance of the energy on the spin appears
when the relativistic terms depending on the velocity of the
particle are taken into account. From the vectors $\vec{s}$,
$\vec{n}$ and $\vec{v}$, a {\it true scalar} can be formed,
$\hat{n}\times\vec{v}.\vec{s}$. The spin-orbit coupling operator is
therefore, 
\beg
V_{sl} = - \phi(r)\hat{n}\times\vec{v}.\vec{s}
\eeq
where $\phi(r)$ is a function of $\vec{r}$. Since
$\vec{l}=\vec{r}\times \vec{p}$,
$m\vec{r}\times\vec{v}=\hbar\vec{l}$, the above interaction can
be written as
\beg
V_{sl} = - \rho(r)\vec{l}.\hat{s}
\eeq
where $f=\hbar\phi/rm$. [This interaction is of first order in
$v/c$ whereas the spin-orbit coupling of an electron in an atom
is a second-order effect. If $V$ is the potential energy, the
atomic spin-orbit interaction,
\beg
{1\over2m^2c^2} {1\over r} {dV\over dr} (\vec{L}.\vec{S})\simeq{1\over
m^2c^2} {V\over a^2} pa\hbar \simeq{v^2\over c^2} V
\eeq
where $a$ represents the linear dimensions of the system and
$\hbar/a \sim p \sim mv$. Therefore, the usual spin-orbit
interaction is of the order of $(v^2/c^2)V$. The interaction (6)
derived from the pseudoscalar should not be confused with the
atomic spin-orbit interaction (7)]. The force being considered
here depends on the spin even in the non-relativistic
approximation, whereas the non-relativistic interaction of
electrons is not dependent on the spin. The energy of the
spin-orbit interaction is mainly concentrated near the surface
of the nucleus, i.e., the function $f(r)$ decreases rapidly
inside the surface. The interaction brings about a splitting of
the level with orbital angular momentum $l$ into two levels with
the angular momentum, $j=l\pm1/2$. Since,
\begin{eqnarray}
-\hat{l}.\hat{s}&=& -{1\over2} l~~ \mbox{for}~~ j=l+{1\over2}\nonumber\\
&=& +{1\over2}(l+1)~~ \mbox{for}~~ j=l-{1\over2}
\end{eqnarray}
the energy difference between the two states is,
\begin{eqnarray}
\Delta E&=&E_{(l-1/2)}-E_{(l+1/2)}\nonumber\\
&=&<f(r)>(l+1/2)\,\,.
\end{eqnarray}
The level with $j=l+1/2$ (the vectors $\vec{l}$ and $\vec{s}$
parallel) is below the level $j=l-1/2$ so that $<f(r)>$ is
positive. As $\l\to\infty$, the lower level at $l+1/2$ goes to
$-\infty$ and the higher level at $l-1/2$ goest to $+\infty$.
Therefore, as $l$ increases, there is a divergence at
$l=\infty$. 

The force on the electron is determined from (6) by using the
relation $\vec{F}=-\partial V/\partial r$. Therefore,
\beg
\vec{F} = {\partial\over\partial r} [f(r)\vec{l}.\vec{s}]
\eeq
is a force on the electron due to pseudoscalar interaction. This
force has not been known previously for electrons in a magnetic
field. It depends on spin and hence is a new force. If a field
is present, the linear momentum $\vec{p}$ is replaced by
$p-(e/c)\vec{A}$ where $\vec{A}$ is the vector potential of the
electromagnetic field. We calculate the contribution of
$-(e/c)A$ to (5) as,
\beg
-\phi(r)\hat{n} \times {\vec{p}\over m}.\vec{s} =
{\phi(r)\over m}\hat{n}\times(e/c)\vec{A}.\vec{s}\,\,.
\eeq
Since the vector potential $\vec{A}=\vec{H}\times\vec{r}$, the
above contribution becomes,
\beg
{\phi(r)e\over cm} {\vec{r}\over|\vec{r}|} \times
\vec{r}.\vec{s} = {\phi(r)e\over cm} \vec{r} \times
(\vec{s}\times \vec{r})\cdot \vec{H}=\vec{\mu}_{B1}\cdots \vec{H}
\eeq
This term is equivalent to appearance of an additional magnetic
momentum whose operator is,
\beg
\mu_{B1} = {\phi(r)e\over cm} \vec{r}\times(\vec{s}\times
\vec{r})\,\,\,. 
\eeq
Thus the interaction of the form (5) produces a correction to
the Bohr magneton.

\section{Effective charge}

We consider the spin-orbit interaction of the type (6) and not
of the type (7) so that the angular momenta combine as
\beg
g_j\vec{j}=g_s\vec{s}+g_l\vec{l} = {1\over2}(g_l+g_s)\vec{j} +
{1\over2} (g_l-g_s)(\vec{l}-\vec{s}\,\,.
\eeq
{}[For conduction electrons $l=0$ and the Lande's
splitting factor is given by
\beg
g = 1+{J(J+1)-L(l+1)-S(S+1)\over2J(J+1)}
\eeq
which is not being considered in the present paper].\\
However, in the present case large values of $l$ can arise.
Multiplying both sides of (14) by $\vec{j}=\vec{l}+\vec{s}$ and
taking eigenvalues, we find,
\beg
g_jj(j+1) = {1\over2}(g_l+g_s)j(j+1) +
{1\over2}(g_l-g_s)[l(l+1)-s(s+1)] 
\eeq
which upon substituting $s={1\over2}$ gives,
\beg
g_j=g_l\pm {g_s-g_l\over2l+1}
\eeq
for $j=l\pm{1\over2}$. For $g_s=2, g_l=1$, we find
\beg
g_\pm = 1 \pm {1\over2l+1}\,\,\,.
\eeq
The cyclotron frequency is defined in terms of the magnetic
field as,
\beg
\omega = {eB\over mc}\,\,\,.
\eeq
Corresponding to this frequency, the voltage along the $y$
direction is,
\beg
\hbar\omega =  eV_y\,\,\,.
\eeq
From (19) and (20),
\beg
{\hbar eB\over mc} = eV_y
\eeq
or
\beg
{e^2B\over2\pi mc} = {e^2\over h} V_y
\eeq
which describes the current in the $x$ direction so that 
\beg
\rho_{xy} = {h\over \nu e^2}\,\,\,.
\eeq
This expression agrees with that of von Klitzing  for $\nu=1$.
We take into account the gyromagnetic ratio from (18) so that
the current (22) may be written as
\beg
I_x = {1\over2} g {e^2B\over 2\pi mc} = {1\over2} g{e^2\over h} V_y\,\,.
\eeq
For $l=0, g=2$,
\beg
I_x = {e^2\over h} V_y
\eeq
which describes the quantized current currectly for $\nu=1$.\\
From (24)
\beg
\nu = {1\over2} g_\pm
\eeq
which gives one value for $+$ sign and the other value for $-$
sign. For $l=0$, we obtain ${1\over2} g_+=1$ and ${1\over2}g_-=0$,
for $l=1, {1\over2}g_+={2\over3}$ and ${1\over2}g_-=1/3$. These
values of $\nu={1\over2}g_\pm$ are given in Table 1 of
Shrivastava [4]. Some of the examples are given in Table 1.
\vskip0.35cm
\begin{center}
Table 1. Predicted values of the fractional charge\,\,.
\vskip0.25cm
\begin{tabular}{ccc}
\hline
$l$ & $g_-/2=l/(2l+1)$ & $g_+/2=(l+1)/(2l+1)$\\
\hline
0 & 0    & 1\\
1 & 1/3  & 2/3\\
2 & 2/5  & 3/5\\
3 & 3/7  & 4/7\\
4 & 4/9  & 5/9\\
5 & 5/11 & 6/11\\
6 & 6/13 & 7/13\\
$\cdots$ & $\cdots$ & $\cdots$\\
$\infty$ & 1/2 & 1/2\\
\hline
\end{tabular}
\end{center}
\vskip0.25cm
\noindent Since
\beg
\hbar\omega_c = g\mu_BB
\eeq
and the Bohr magneton, $\mu_B=e\hbar/2mc$, it may be agreed that
the effective charge is,
\beg
e_{eff} = ({1\over2}) ge = \nu e.
\eeq
Therefore for $\nu=1/3$, the effective charge becomes $(1/3)e$.
All of the predicted values given in Table 1 are exactly the
same as those found in experimental measurements [6-10]. It may
be noted that when we fix the magnetic field at $\nu=1/3$, all
of the electrons point to spin $=-1/2$ so that the system
resembles a ferromagnet. At $\nu=1/2$, the value is approached
as a limit of $l\to\infty$ from both the spin configurations so
that the fluid is a singlet with one series having spin
$-{1\over2}$ and the other having spin $+{1\over2}$. So far, we
have not used the concept of Landau levels. If we introduce the
Landau levels, then the above angular momenta should occur in
each and every Landau level. Assuming that $n$ is the Landau
level quantum number, the effective charge of $n\nu$ becomes
observable. The predicted limiting value of 1/2 becomes $n/2$
and all of the fractions, 1/2, 2/2, 3/2, 4/2, 5/2, 6/2 and 7/2, etc. 
become observable. This predicted result is the same as
experimentally found by Yeh et. al [11].
 The experimentally measured values of
the Hall resistance are given by St\"ormer [12]. Starting from the high field side, the
fractions 1/3, 2/5, 3/7, 4/9 and 1/2 are the same as predicted
by column 2 of Table 1. Similarly on the left hand side 2/3,
3/5, 4/7 are the same as the 3rd column of Table 1. Due to
Landau level quantum number, $n$, the fractions $n\nu$ are also
observable. Therefore, $2\times 2/5=4/5, 2\times 2/3=4/3$ as
well as $n/3=5/3$ and $n$ times 1 which is 1,2,3 and 4 are all
clearly observed.

\section{Conclusions}

We have derived an interaction for the electrons in a magnetic field
which appears like a spin-orbit interaction and  successfully explains
all of the observed plateaus found experimentally in the quantum Hall
effect. It predicts a correction to the value of the Bohr magneton.
\newpage
\noindent{\bf References}
\begin{enumerate}
\item K.v. Klitzing, G. Dorda and M. Pepper, Phys. Rev. Lett.
	{\bf45} (1980) 494.
\item D.C. Tsui, H.L. St\"ormer and A.C. Gossard, Phys. Rev.
	Lett. {\bf48} (1982) 1559.
\item R.B. Laughlin, Phys. Rev. Lett. {\bf50} (1983) 1395.
\item K.N. Shrivastava, Phys. Lett. A{\bf113} (1986) 435;
	{\bf115} (1986) 459(E).
\item K.N. Shrivastava, Mod. Phys. Lett. {\bf13} (1999) 1087.
\item K.N. Shrivastava, Superconductivity: Elementary Topics,
	World Scientific, Singapore 2000.
\item R. Willett, J.P. Eisenstein, H.L. St\"ormer, D.C. Tsui,
	A.C. Gossard and J.H. English, Phys. Rev. Lett. {\bf59}
	(1987) 1776. 
\item J.P. Eisenstein and H.L. St\"ormer, Science {\bf 248}
	(1990) 1510.
\item R.R. Du, H.L. St\"ormer, D.C. Tsui, L.N. Pfeiffer and K.W.
	West, Phys. Rev. Lett. {\bf70} (1993) 2944.
\item W. Pan, J.S. Xia, V. Shvarts, D.E. Adams, H.L. St\"ormer,
	D.C. Tsui, L.N. Pfeiffer, K.W. Baldwin and K.W. West,
	Phys. Rev. Lett. {\bf 83} (1999) 3530.
\item A.S. Yeh, H.L. St\"ormer, D.C. Tsui, L.N. Pfeiffer, K.W.
	Baldwin and K.W. West, Phys. Rev. Lett. {\bf62} (1999)
	592. 
\item H.L. St\"ormer, Rev. Mod. Phys. {\bf71} (1999) 875.
\end{enumerate}

\end{document}